\documentclass[a4paper,11pt]{article}
\usepackage{pos}

\title{Machine learning approaches to the QCD transition}

\author*[a,b,c]{Andrea Palermo}
\author[a,b]{Lucio Anderlini}
\author[b]{Maria Paola Lombardo}
\author[d]{Andrey Kotov}
\author[e]{Anton Trunin}

\affiliation[a]{Dipartimento di Fisica e Astronomia, Universit\`a di Firenze,\\
  Via G. Sansone 1, Sesto Fiorentino I-50019, Firenze, Italy}

\affiliation[b]{INFN sezione di Firenze,\\
Via G. Sansone 1, Sesto Fiorentino I-50019, Firenze, Italy}

\affiliation[c]{Institut für Theoretische Physik, Johann Wolfgang Goethe-Universität, \\Max-von-Laue-str. 1, 60438 Frankfurt am Main, Germany}

\affiliation[d]{Bogoliubov Laboratory of Theoretical Physics (BLTP), Joint Institute for Nuclear Research (JINR),\\ Dubna, 141980 Russia}

\affiliation[e]{Samara National Research University,\\ Samara, 443086 Russia}

\emailAdd{andrea.palermo@fi.infn.it}
\emailAdd{Lucio.Anderlini@cern.ch}
\emailAdd{lombardo@fi.infn.it}
\emailAdd{Kotov.andrey.yu@gmail.com} 
\emailAdd{amtrnn@gmail.com}

\abstract{We study the high temperature transition in pure $SU(3)$ gauge theory and in full QCD with 3D-convolutional neural networks trained as parts of either unsupervised or semi-supervised learning problems. Pure gauge configurations are obtained with the MILC public code and full QCD are from simulations of $N_f=2+1+1$ Wilson fermions at maximal twist. We discuss the capability of different approaches to identify different phases using as input the configurations of Polyakov loops. To better expose fluctuations, a standardized version of Polyakov loops is also considered.}

\FullConference{%
 The 38th International Symposium on Lattice Field Theory, LATTICE2021
  26th-30th July, 2021
  Zoom/Gather@Massachusetts Institute of Technology
}

\begin{document}
\maketitle

\section{Introduction}
A physical system exhibits different macroscopical behaviors depending on the thermodynamic parameters describing its equilibrium. Such behaviors distinguish the \emph{phases of matter}, which may change undergoing a \emph{phase transition}. Usually, the phases of a system can be classified by some \emph{order parameter}, the value of which indicates the phase of the system. Typical examples of order parameters are the magnetization in the Ising model or the Polyakov loop in the SU(3) gauge theory. On the other hand, the definition of an order parameter is not always possible or straightforward: a typical example is confinement in QCD. The standard lore associates confinement with chiral symmetry breaking, or, more generically, assumes that confinement would imply chiral breaking; however, it has also been hypothesized that confinement persists up to temperature $T_d$ of about twice the chiral transition temperature $T_\chi$. If this is the case, a new phase would appear between $T_\chi$ and $T_d$. Machine learning techniques may succeed in identifying such a phase in a model independent way.

Nowadays, machine learning is applied successfully in many areas of physics. A common problem where neural networks are employed is classification, where the network is trained to distinguish different input data according to some features of the input itself. Since the distinction of phases of matter can be regarded as a classification problem, machine-learning algorithms might help in classifying phases of matter even when a properly-said order parameter is lacking. 

 There have been several other attempts to use neural networks or AI methods to classify phases and identify phase transitions \cite{Giannetti:2018vif}. So far, these studies have been mostly restricted to spin models, see e.g. Ref.\cite{Cole:2020hjx, Alexandrou:2019hgt} for a small subset of available studies, which contains a more complete set of references. Analysis of the thermal transition in  Yang-Mills have appeared \cite{Boyda:2020nfh,Wetzel:2017ooo}, but the investigations of gauge models, and of the even more complex fermion-gauge models such as QCD are scarce.

In this study, we probe the capabilities of a neural network to distinguish among different phases of quantum field theories in an unsupervised scheme, that is without using predefined labels for different configurations during the training of the network.  We consider pure gauge as well as full QCD. SU(3) and QCD configurations are obtained from lattice simulations in a 3+1 dimensional lattice.
 We deal with the complications of the four dimensional gauge dynamics by defining the three dimensional Polyakov loop configurations: at each space point and on  each configuration we compute the Polyakov loop, which is then used as input for the unsupervised classification problem.

\section{Phase transitions as unsupervised learning problems}
\begin{figure}
    \centering
    \includegraphics{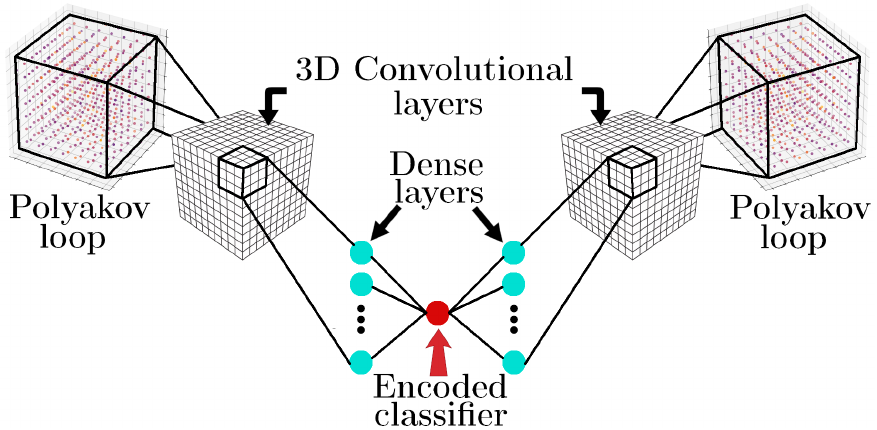}
    \caption{Scheme of the autoencoder used in this study. In the encoder, the Polyakov loop configuration is fed into 3D convolutional and then to dense layers; thereby the encoded classifier is obtained. A similar architecture is used for the decoder, which starting from the encoded classifier reconstructs the input configuration.}
    \label{figurina autoencoder}
\end{figure}

To classify configurations of Polyakov loops at different temperatures, we build 3D-convolutional autoencoders \cite{autoencoder} using TensorFlow and Keras \cite{repository,tensorflow,keras}. An autoencoder is a compound of two neural networks: an encoder that condenses the input information, for example to
a single number, and a decoder, which reconstructs the input data from the compressed ones. Here, the
encoder processes the information contained in the Polyakov loops configuration to a single number,
named \emph{encoded classifier} (see figure \ref{figurina autoencoder}).

The autoencoder is trained, as a whole, to reproduce as output its own input. When this is achieved,
the encoded classifier effectively \emph{encodes} the most important feature(s) describing the variety of the
input. The mapping of the input to the encoded classifier, however, can be quite complicated. To simplify matters, one can perform a \emph{semi-supervised} training by pinning some of the input configurations at extreme temperatures to predefined values of the encoded classifier.
In such a scheme, unlabelled configurations similar to those pinned somewhere in the latent
space are clustered together, and the interpretation of the encoded classifier is easier.
Assuming lattice configurations simulated at different temperatures are mainly distinguished by their
degree of disorder, the encoded classifier may provide an effective order parameter for an arbitrary lattice
configuration, independently of the underlying theory.

\section{Data-set}
For this study, we have used configurations from different lattices depending on the theory. For the pure $SU(3)$ gauge theory we used $8^3\times4$ lattice configurations generated using the MILC public code \cite{milc}.  For this geometry and action, the pseudocritical coupling is  $\beta_c = 5.69(2)$,  giving a  the critical temperature $T_c\sim 260 \ {\rm MeV}$ \cite{Boyd:1996bx}. We analysed $30$ configurations of Polyakov loops for each temperature. The configurations span a wide range of the coupling parameter, from strong to very weak coupling; we formally express the results as a function of $T/T_c$  for
convenience of comparison with dynamical studies, and of course the largest
values have a limited meaning - they were just meant to probe the system as close as possible to the free regime. 

The full QCD configurations are obtained from simulations of $N_f=2+1+1$ Wilson fermions at maximal twist on a lattice of $32^3$ space dimension \cite{twist}.
 The strange and charm masses have their physical values, while the pion mass is 370 MeV. 
 This relatively large value helps the analysis with the Polyakov loops, and among the various future step of interest there is of course the study of the behaviour closer to the chiral limit. The pseudocritical temperature is $T_c\sim 200 {\rm \ MeV}$. For each temperature studied, $200$ Polyakov loops configurations have been used.
 
In both cases, for the training set a fraction $60\%$ of the configurations is randomly selected, and the remaining configurations define the validation set. Only the training set is used by the neural network in the training process. In this paper, we use the known value of the critical temperatures only to highlight visually in the figures where we expect the transition to happen. Such information is not available to the neural network. 

\section{Results}
\begin{figure}
     \centering
     \includegraphics[width=.4\textwidth,,trim=0 16 0 0, clip]{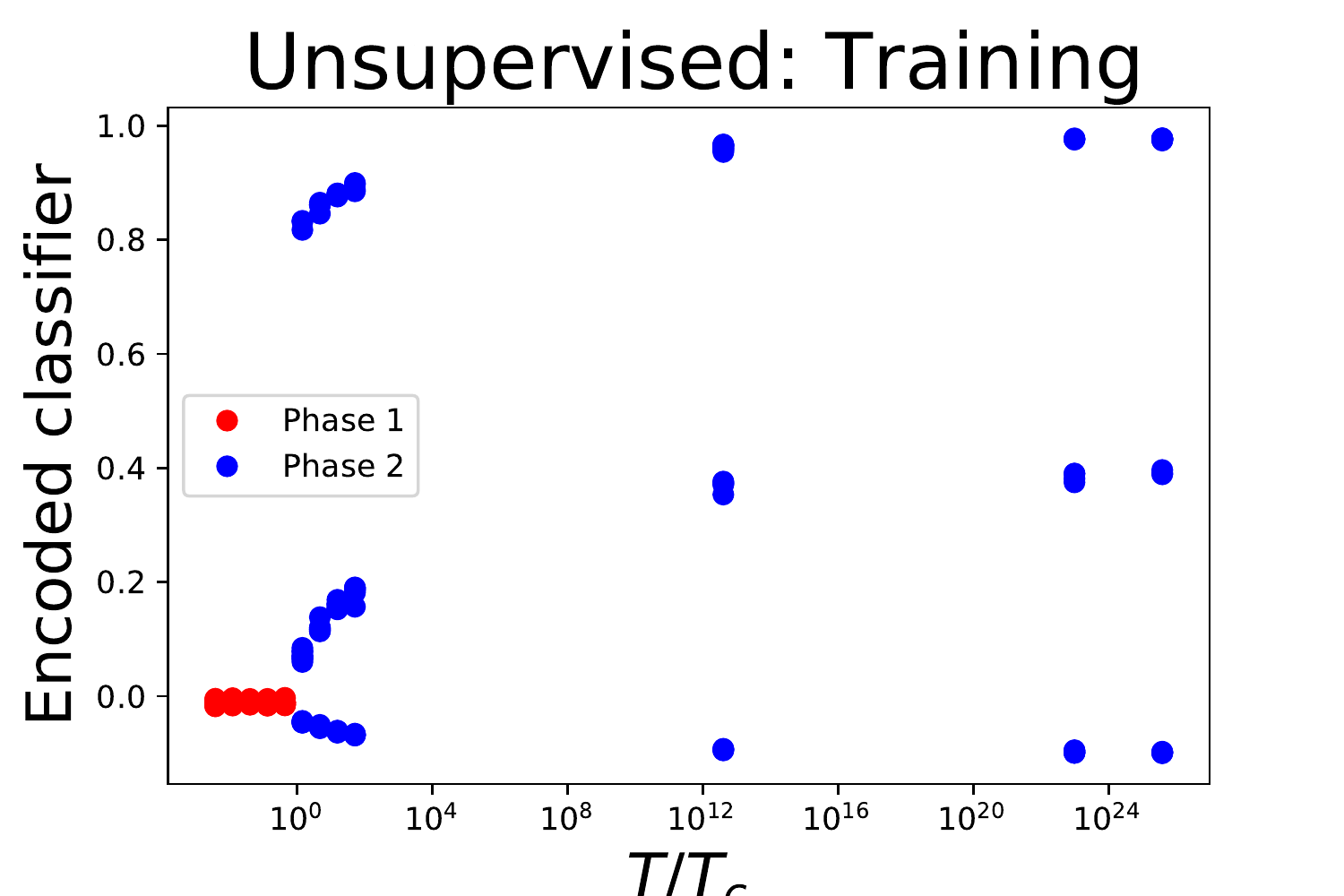}
     \includegraphics[width=.4\textwidth,trim=0 16 0 0, clip]{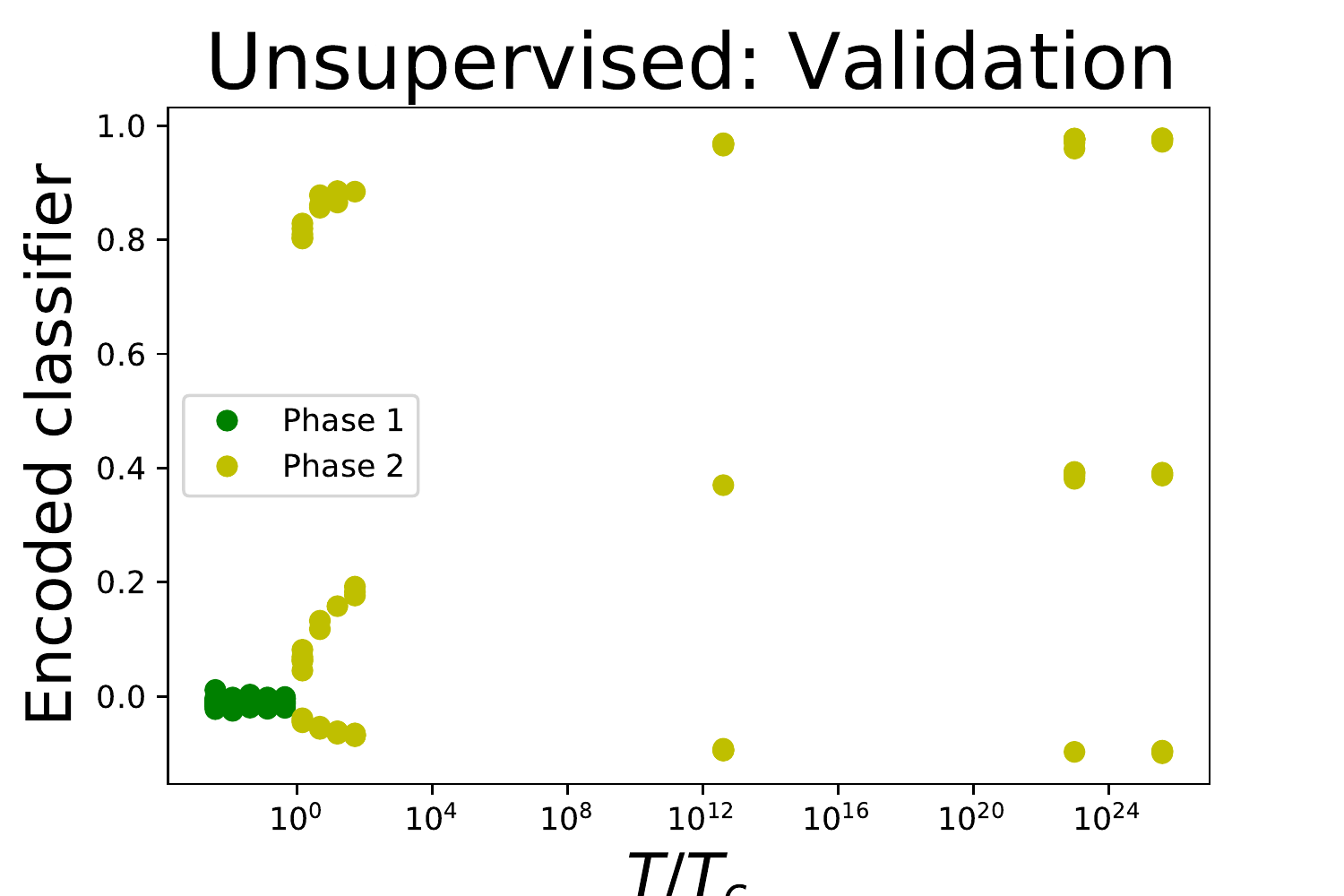}
     \put(-90,-8){$T/T_c$}
	 \put(-265,-8){$T/T_c$}
     \caption{The encoded classifier for the training and validation sets of Polyakov loop configurations in the unsupervised scheme. A change in the behavior of the encoded classifier is visible at $T=T_c$. Three values of the encoded classifier are equally possible for temperature higher than $T_c$, which can be interpreted as the spontaneous symmetry breaking of the centre symmetry $\mathbb{Z}_3$.}
     \label{fig:unsupervised gauge}
 \end{figure}

In the case of pure $SU(3)$ gauge theory, the mean Polyakov loop is an exact order parameter for confinement  in an infinite volume.
A machine learning approach to finite size scaling 
in a spin model may be found in Ref.\cite{Kim:2021ulz}.
Training the autoencoder as an unsupervised and semi-supervised classification problem we obtain an encoded classifier clearly related to the order parameter.
 Indeed, two classes are identified by the encoded classifier below and above $T_c$. The unsupervised scheme highlights the $\mathbb{Z}_3$ symmetry breaking: three different values of the encoded classifier are equally possible for the gauge theory at temperature higher than $T_c$, whereas for $T<T_c$ there is only one possibility (figure \ref{fig:unsupervised gauge}).  

\begin{figure}
     \centering
     \includegraphics[width=.4\textwidth,,trim=0 16 0 0, clip]{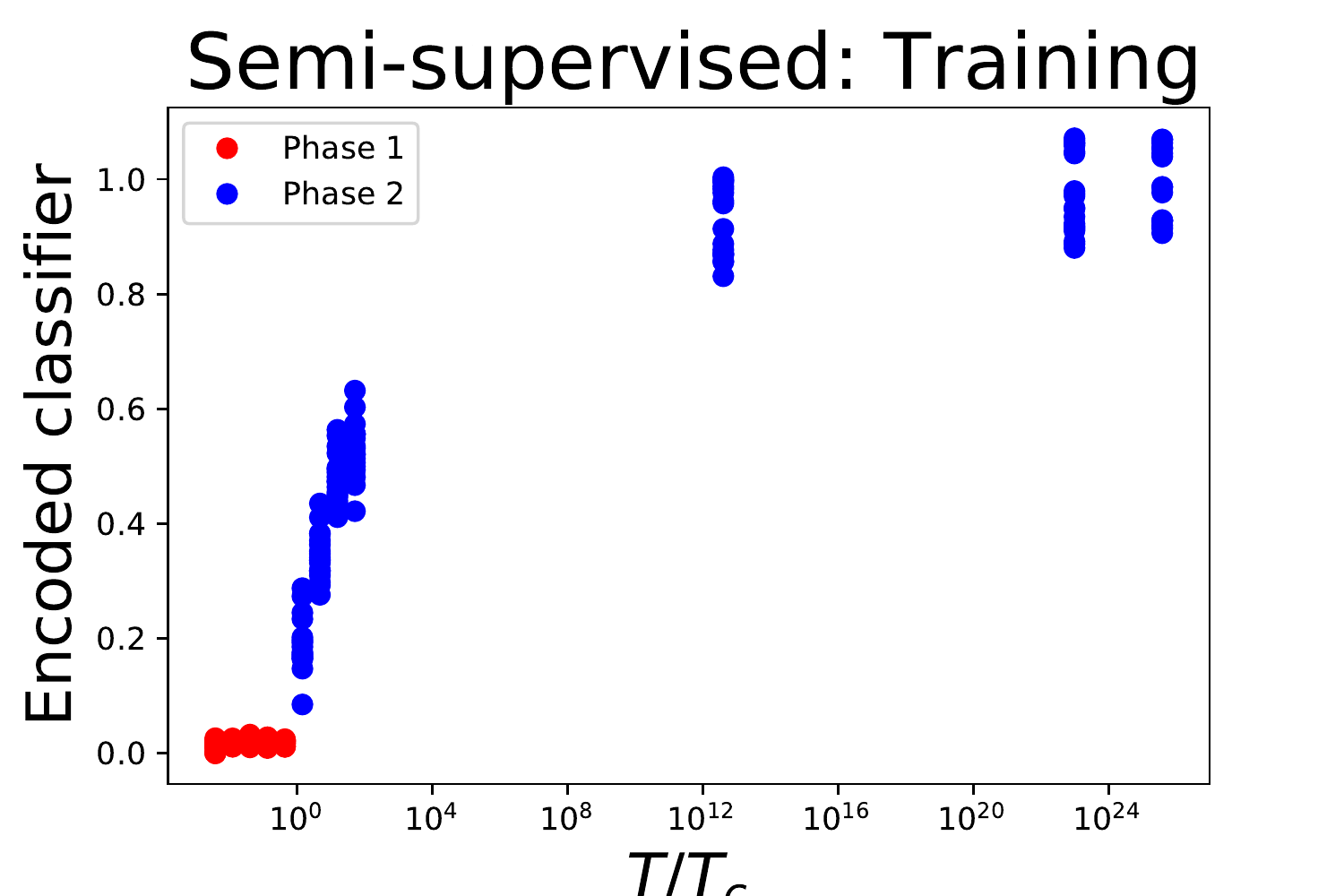}
     \includegraphics[width=.4\textwidth,trim=0 16 0 0, clip]{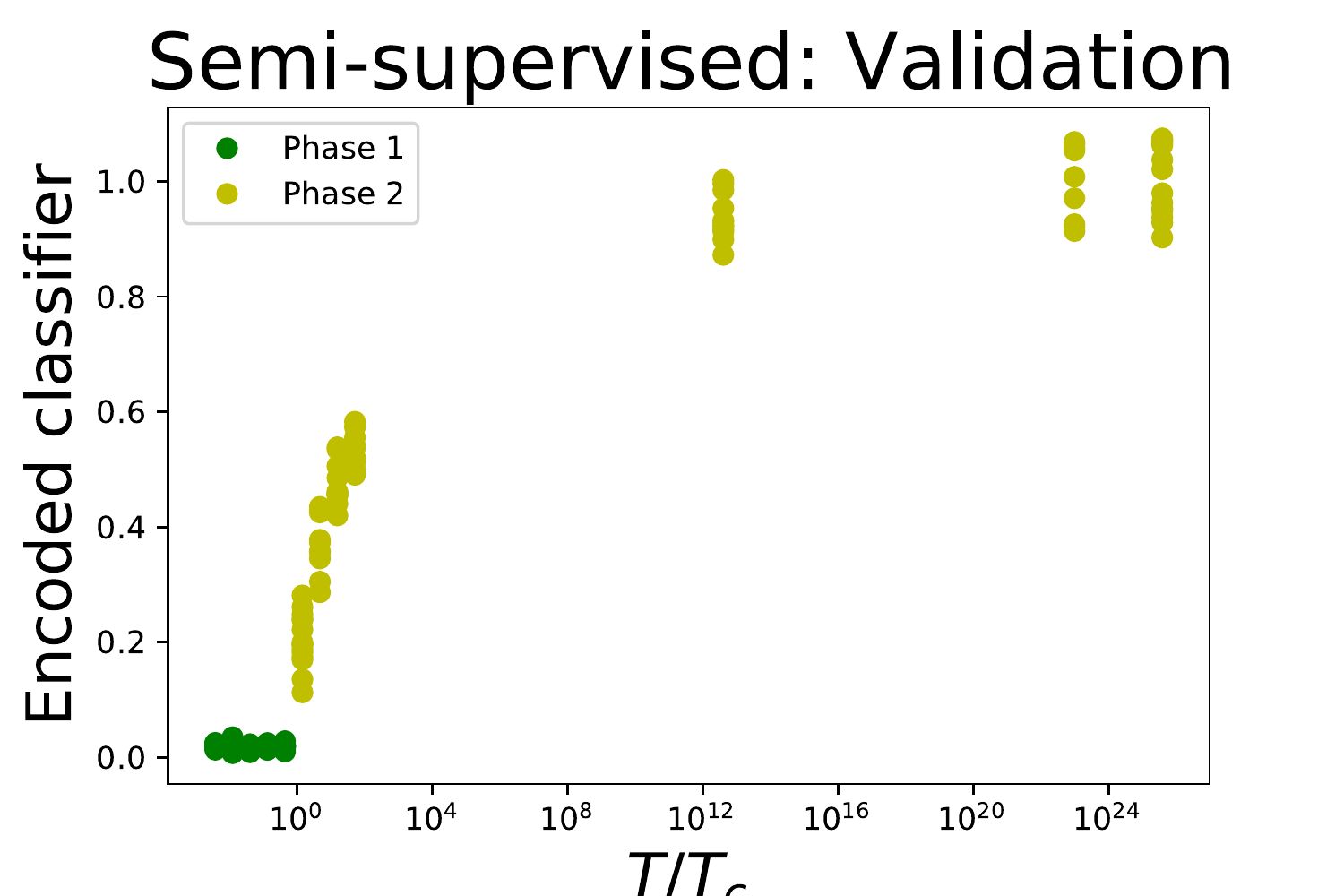}
    \put(-90,-8){$T/T_c$}
	 \put(-265,-8){$T/T_c$}
     \caption{The encoded classifier for the training and validation sets of Polyakov loop configurations in the semi-supervised scheme. The algorithm recognises configurations below $T_c$ as similar to those whose encoded classifier has been set to $0$, and for very high temperature they are clustered around $1$. The change in behavior around $T_c$ is clearly visible.}
     \label{fig:semisupervised gauge}
 \end{figure}

  For the semi-supervised learning problem, we pin a fraction of $\sim20\%$ of the training configurations at the lowest and highest values of $T/T_c$ to predefined values of the encoder classifier, in this case $0$ and $1$ respectively\footnote{Neither the latent space nor the encoded classifier have a physical meaning, so that we can use arbitrary numbers for this pinning.}. This procedure strengthens the correlation of the encoded classifier with the true order parameter (fig. \ref{fig:semisupervised gauge}).
 
   \begin{figure}
     \centering
     \includegraphics[width=.4\textwidth,,trim=0 16 0 0, clip]{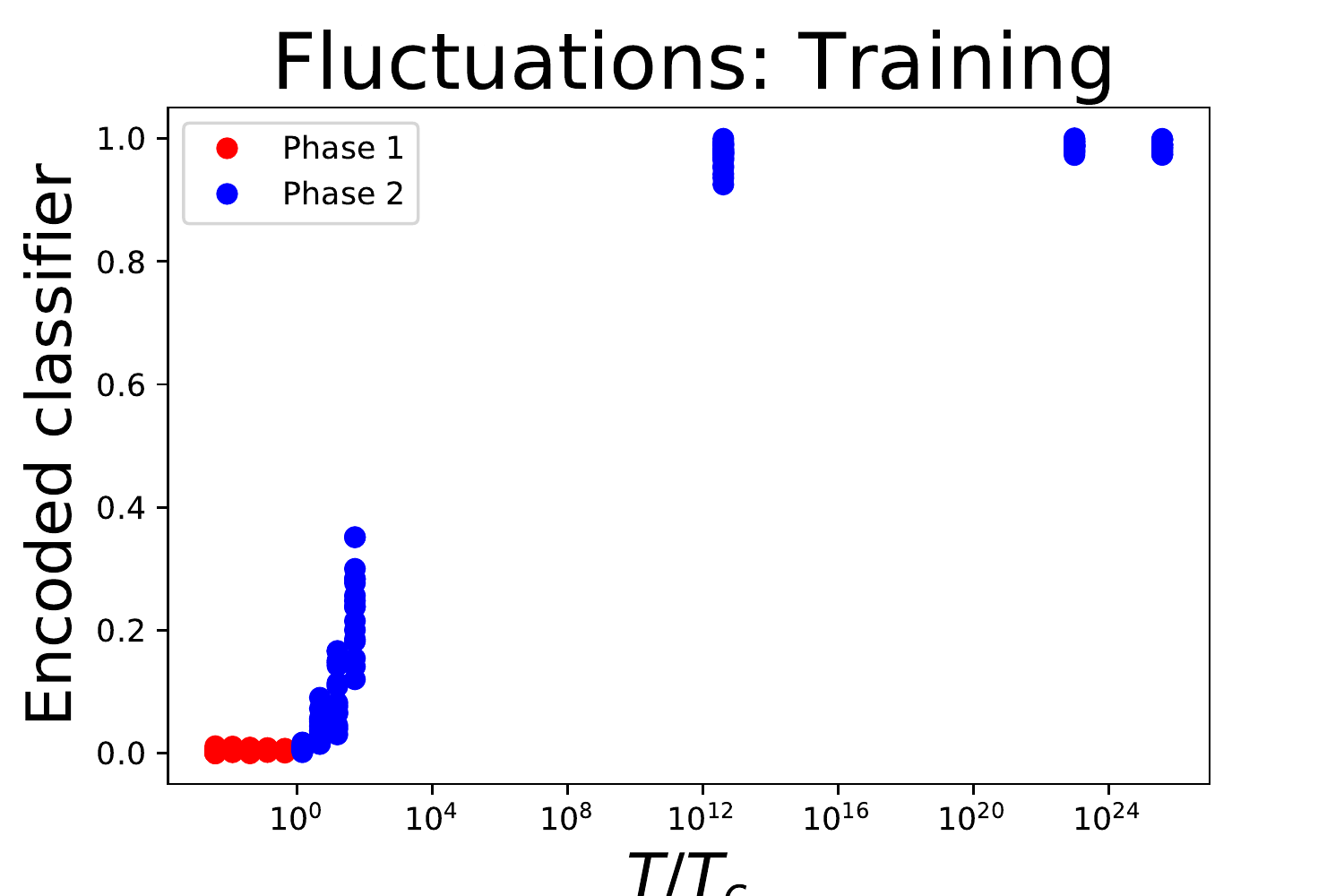}
     \includegraphics[width=.4\textwidth,trim=0 16 0 0, clip]{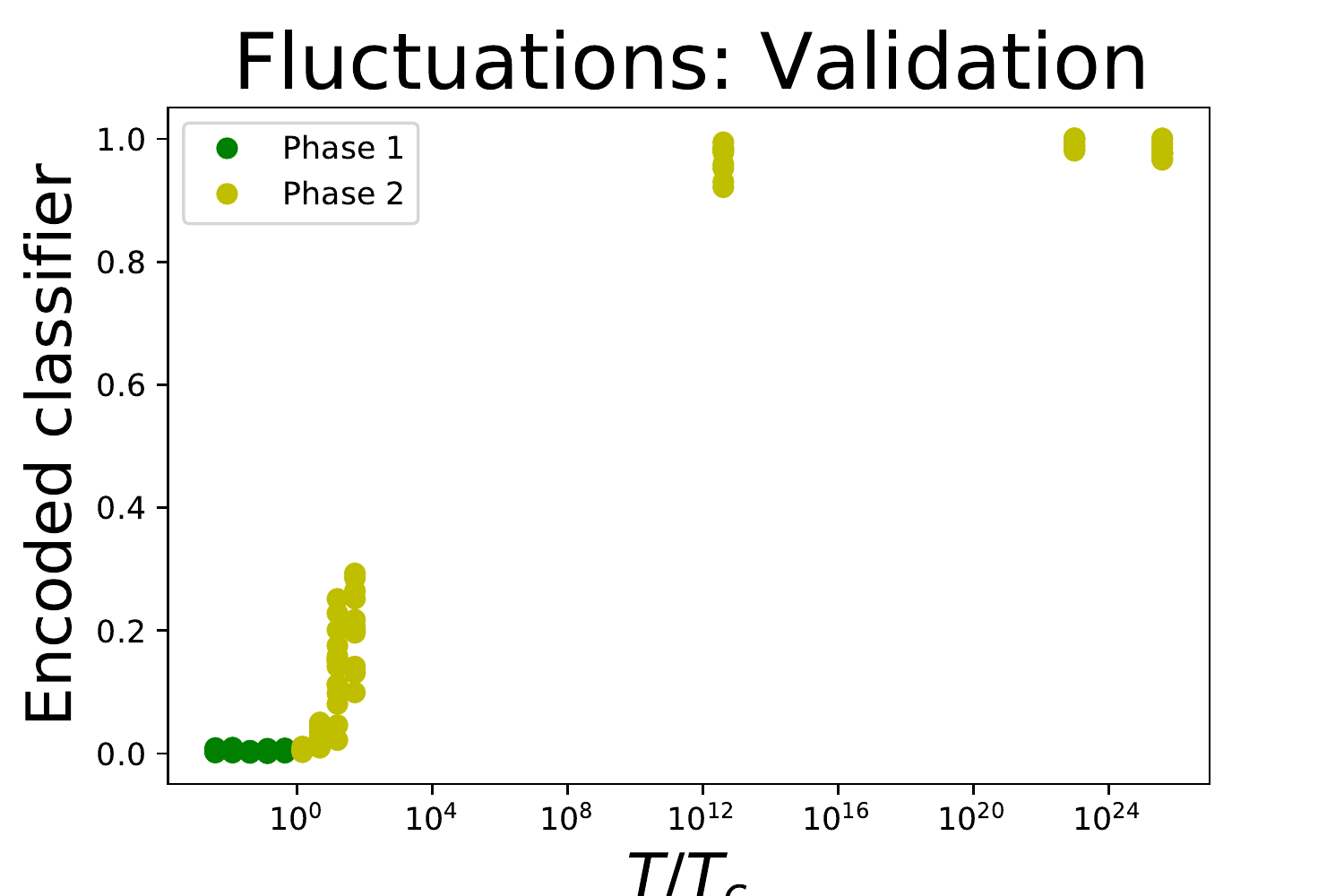}
     \put(-90,-8){$T/T_c$}
	 \put(-265,-8){$T/T_c$}
     \caption{The encoded classifier for the training and validation sets of the standardized Polyakov loop in the semi-supervised scheme. Configurations at $T\sim T_c$ are more difficult to discern, but the change in the encoded classifier is still visible.}
     \label{fig:gauge fluct}
 \end{figure}
 
  In order to ensure that the network is learning from the texture on the lattice rather than simply averaging the Polyakov loops on each simulation, we define a standardized Polyakov loop. Let $P^{(i)}$ be the $i$-th 3d Polyakov loop configuration and $\mu_i$ the mean value of $P^{(i)}$. Denoting with the indices $x,y,z$ the spatial coordinates on the lattice, running from $1$ to the lattice dimension $L$, we define a standard deviation: $$\sigma_i=\sqrt{\frac{1}{2\times L^3}\sum_{x,y,z=1}^{L}|{P^{(i)}}_{xyz}-\mu_i|^2}.$$ 
  The components of the $i$-th standardized configuration are:
 $$\tilde{P}^{(i)}_{xyz}=(P^{(i)}_{xyz}-\mu_i)/\sigma_i.$$
  Despite a slight loss in precision, the network is still perfectly able to identify the presence of a phase transition at $T \sim T_c$ even when using the standardised Polyakov loop as input for the network (fig \ref{fig:gauge fluct}).
 
 In the case of QCD, the Polyakov loop is no longer an order parameter and the identification of a phase transition based on the Polyakov loop is
 not theoretically justified. On top of that, the phase transition for the configurations studied is known to be a crossover, so that the change in behavior of the encoded classifier is expected to be milder compared with the $SU(3)$ case.  
 Identifying the pseudocritical temperature is then a significantly different challenge with respect to the pure gauge study. 

\begin{figure}
     \centering
     \includegraphics[width=.4\textwidth,,trim=0 16 0 0, clip]{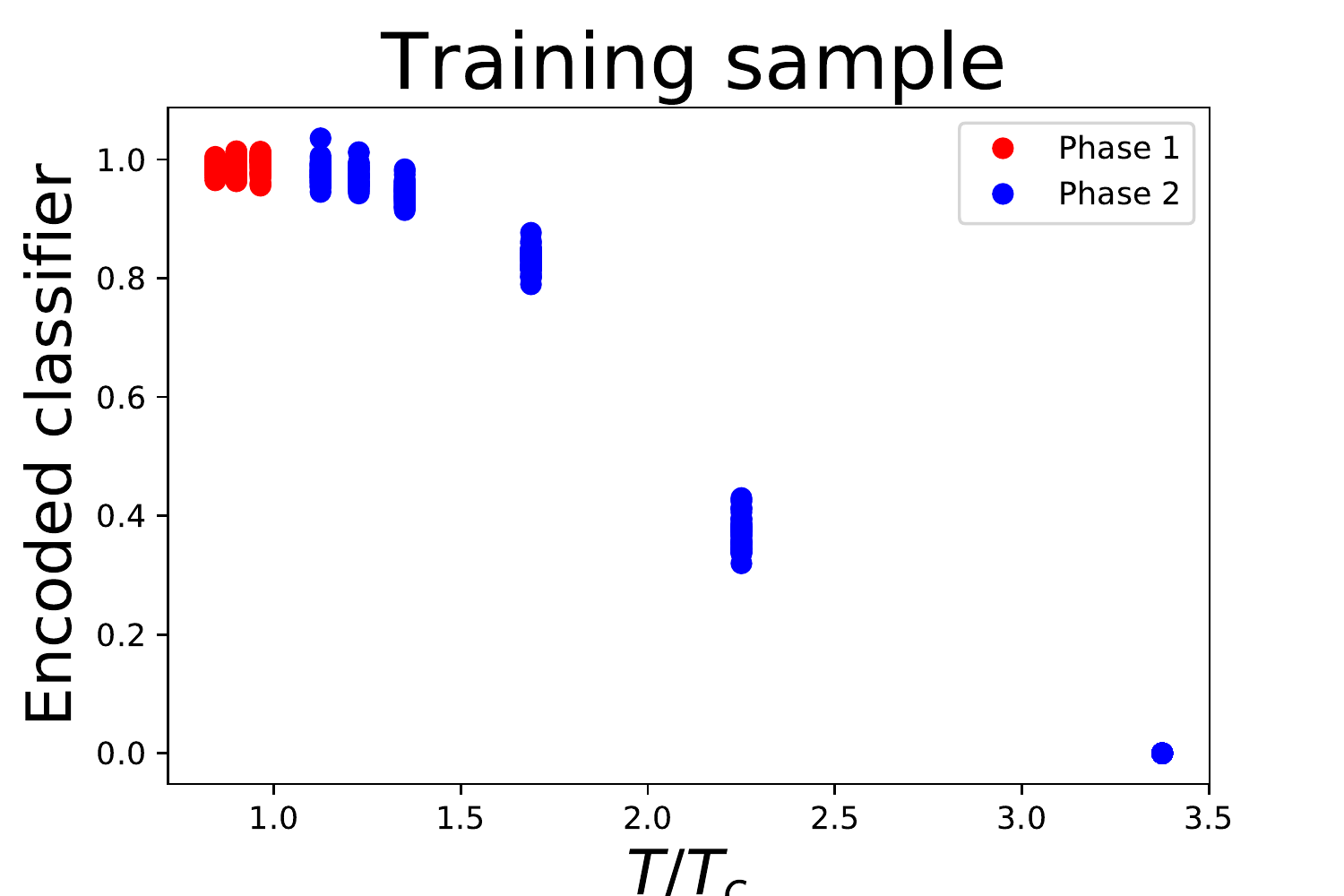}
     \includegraphics[width=.4\textwidth,trim=0 16 0 0, clip]{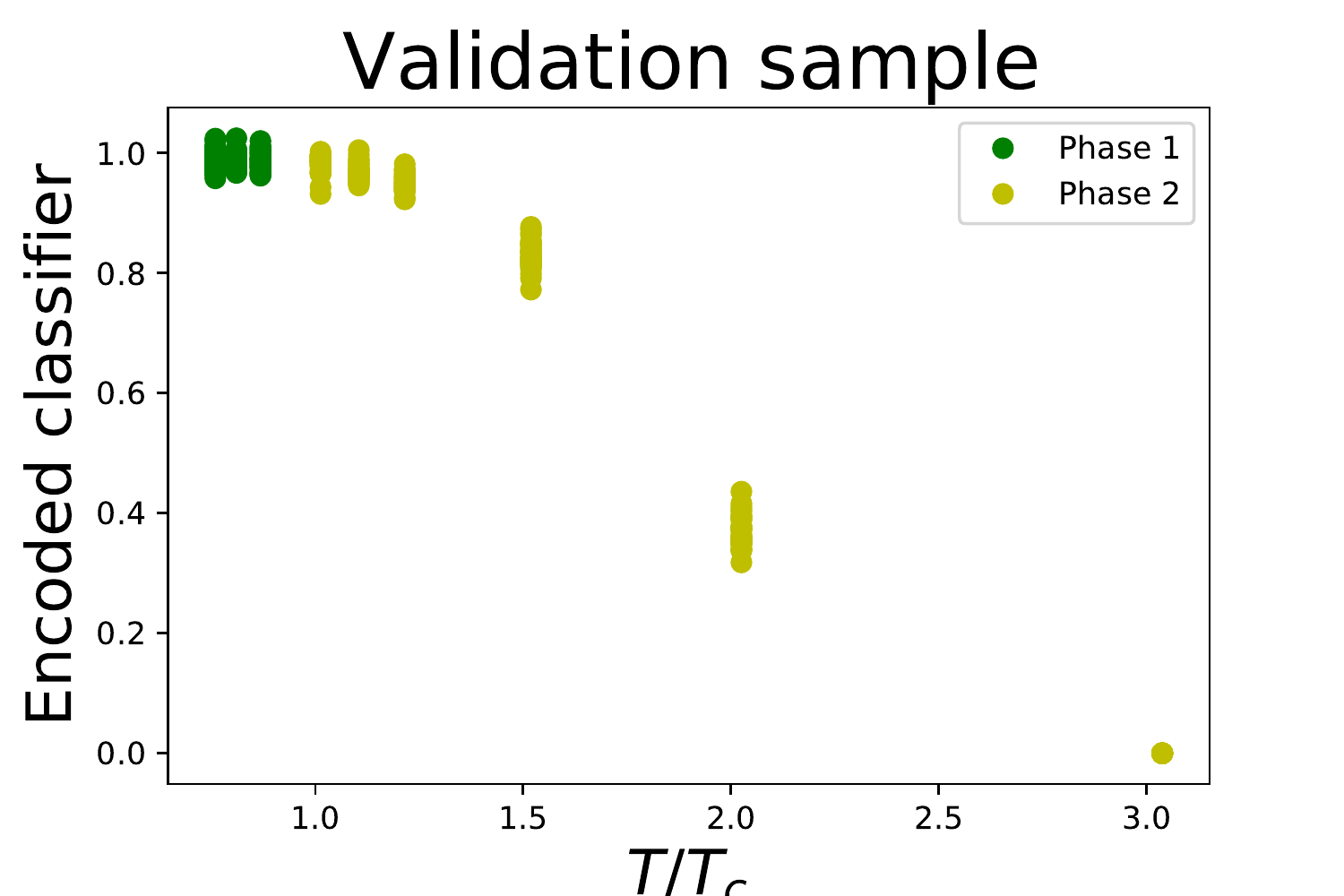}
     \put(-90,-8){$T/T_c$}
	 \put(-265,-8){$T/T_c$}
     \caption{The encoded classifier for QCD configurations in the semi-supervised scheme. The encoded classifier changes smoothly as a function of temperature. Two classes may be identified separated by the temperature $T\sim 1.5T_c$. }
     \label{fig:qcd semi}
 \end{figure}

We study the semi-supervised problem with the Polyakov loop and its standardized version. This time, we pin the configurations at low temperatures to an encoded classifier equal to $1$, and higher temperatures are pinned to $0$. In this case, the encoded classifier turns out to be a much smoother function compared to the pure gauge theory, in agreement with our anticipations. From the encoded classifier, one could identify two classes, separated at temperature $T\sim1.5 T_c$, as shown in figure \ref{fig:qcd semi}. This is a good achievement considering the simple architecture of the neural network used compared to the complexity of the problem. 

\begin{figure}
     \centering
     \includegraphics[width=.4\textwidth,,trim=0 16 0 0, clip]{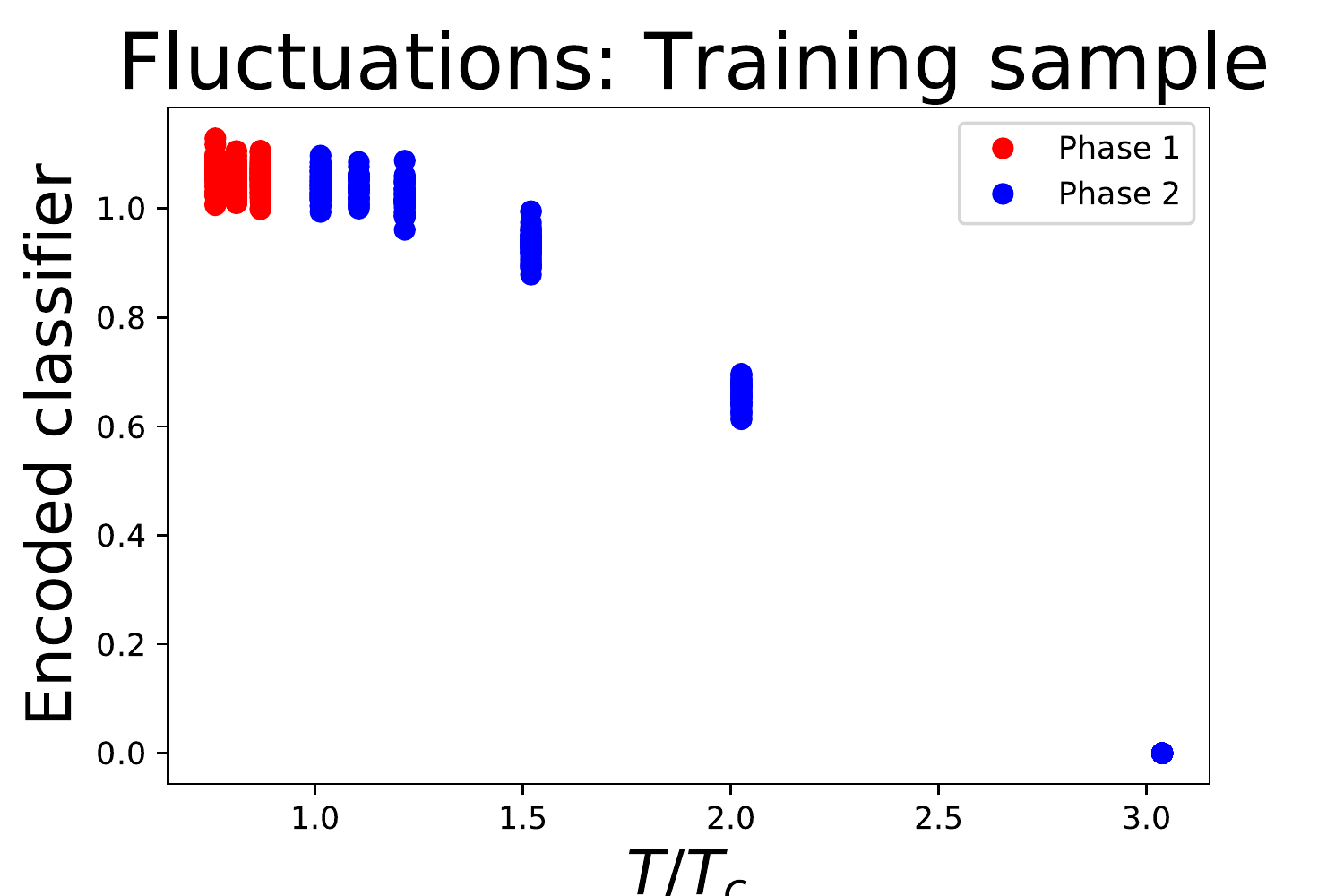}
     \includegraphics[width=.4\textwidth,trim=0 16 0 0, clip]{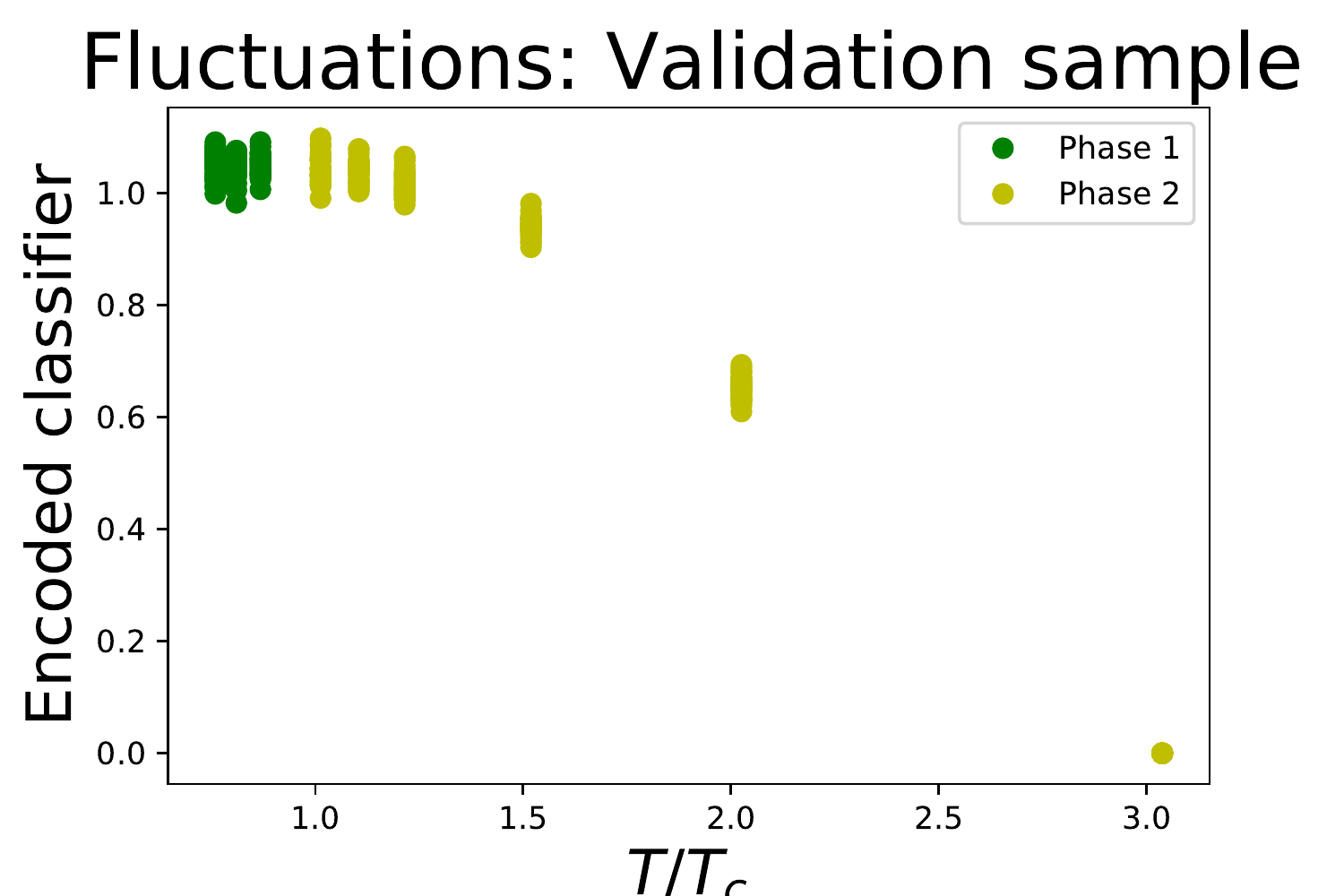}
     \put(-90,-8){$T/T_c$}
	 \put(-265,-8){$T/T_c$}
     \caption{Same as figure \ref{fig:qcd semi}, but the standardized Polyakov loop is used as input. The distinction of two classes is still possible at temperature $T\sim1.5T_c$.}
     \label{fig:qcd fluct}
 \end{figure}

For the standardized configuration, a discussion similar to the previous one can be repeated. Surprisingly, in this case, the loss in precision with respect to the pure Polyakov loop configurations is much lower compared to the $SU(3)$ case, and one could still identify two classes for a critical temperature $T\sim 1.5T_c$ (figure \ref{fig:qcd fluct}). A finer temperature scan could further probe the difference between the two preprocessing.

\section{Conclusions}
We probed the capability of Convolutional Neural Networks trained as either unsupervised or semi-supervised classifiers to identify different phases of gauge theories. We observe a crossover between
the two phases at the expected temperature in a pure gauge theory and a qualitatively similar behavior in
full QCD.  A finer temperature scan, finite-size scaling and continuum limit will improve the performance
of the autoencoder, hopefully providing further insight into ML approaches to the study of phase transitions.

\section{Acknowledgement}
This publication is part of a project that has received funding from the European Union’s Horizon 2020 research and innovation programme under grant agreement STRONG – 2020 - No 824093.

\end{document}